%% file: EuMW_ODDM_multitx_calib.tex
\newcommand{\CLASSINPUTtoptextmargin}{19mm}%
\newcommand{\CLASSINPUTbottomtextmargin}{43mm}%
\newcommand{\CLASSINPUTinnersidemargin}{12.9mm}%
\newcommand{\CLASSINPUToutersidemargin}{12.9mm}%
\documentclass[conference,10pt,a4paper]{IEEEtran}
\newif\ifrelease
\releasetrue 

\ifrelease
\else
	\usepackage[inline]{showlabels} 
\fi

\usepackage{amsmath,amssymb,amsfonts,siunitx, mathtools}
\usepackage{bm}
\usepackage{upgreek}
\usepackage{cancel}
\usepackage{yhmath} 
\usepackage{xspace} 
\usepackage{makecell}
\usepackage{tikz}
\usetikzlibrary{tikzmark,positioning,fit,calc,trees,mindmap}
\usetikzlibrary{datavisualization.formats.functions}
\usepackage{pgfplots}
\usetikzlibrary{patterns}
\usetikzlibrary{shapes.multipart}
\usepackage{subcaption}

\usepackage{times}
\usepackage{graphicx}
\usepackage{multirow}
\usepackage[none]{hyphenat}
\usepackage{float}
\usepackage{subfig}
\usepackage{iftex}

\usepackage[nohyperlinks, nolist]{acronym}
\usepackage{glossaries-extra}
\ifPDFTeX%
\usepackage{t1enc}
\usepackage{times}
\fi%
\ifXeTeX%
\usepackage{fontspec}
\fi%
\ifLuaTeX%
\usepackage{fontspec}
\fi%
\input{EuMW_modify_IEEEtran_18b_CTAN_V5}

\input{commands}
\makeglossaries
\loadglsentries{glossary}

\begin{document}
\input{acronyms}
\raggedbottom
%
%
%
\title{Relaxed Multi-Tx DDM Online Calibration}
%
%
\author{%
\IEEEauthorblockN{%
Mayeul Jeannin\EuMWauthorrefmark{\#1}, 
Oliver Lang\EuMWauthorrefmark{*2},
Farhan Bin Khalid\EuMWauthorrefmark{\#},
Dian Tresna Nugraha\EuMWauthorrefmark{\dag},
and Mario Huemer\EuMWauthorrefmark{*}
}
\IEEEauthorblockA{%
\EuMWauthorrefmark{\#}Infineon Technologies AG, Germany\\
\EuMWauthorrefmark{*}Institute of Signal Processing, Johannes Kepler University Linz, Austria\\
\EuMWauthorrefmark{\dag}PT. Infineon Technologies, Indonesia\\
\EuMWauthorrefmark{1}mayeul.jeannin@infineon.com, \EuMWauthorrefmark{2}oliver.lang@jku.at\\
}
}
%
\maketitle
%
%
%
\begin{abstract}
	In \ac{mimo} radar systems based on \ac{ddm}, \pss are employed in the transmit paths and require calibration strategies to maintain optimal performance all along the radar system's life cycle.
	In this paper, we propose a novel family of \ac{ddm} codes that enable an online calibration of the \pss that scale realistically to any number of simultaneously activated \ac{tx}-channels during the calibration frames.
	To achieve this goal we employ the previously developed \ac{oddm} sequences to design calibration \ac{ddm} codes with reduced inter-\ac{tx} leakage.
	The proposed calibration sequence is applied to an automotive radar data set modulated with erroneous \pss.
\end{abstract}
%
%
\acresetall 

\section{Introduction}

Preserving the radar system's \ac{dr} in the angle dimension is of utmost importance in automotive applications \cite{schmid2013effects}.
The \ac{mimo} concept is largely employed to increase the angular resolution and \ac{dr} of \ac{fmcw} radar systems \cite{fishler2004mimo}.
With more and more \ac{tx} antennas employed, advanced \ac{tx} multiplexing techniques are needed \cite{waldschmidt2021automotive}.
The popular \ac{ddm} technique adds a \ac{tx}-dependent linear phase shift from ramp to ramp such that the signal reflected from a given \ac{tx} appears shifted in the Doppler space \cite{rabideau2011doppler,jansen2019automotive}.

In previous studies, it was shown that the non-idealities of the \ac{tx} \pss reduces the sensitivity of the system as the appearance of spurs reduces the sensitivity on the Doppler and angular spectra \cite{jeannin2022modeling}.
The \pss can be calibrated during radar operation by employing so-called online calibration methods. These methods typically exploit targets of opportunity in the radar's field of view.
In \cite{jeannin2022iterative}, an online calibration method is described that activates only a single \ac{tx}-channel at a time, estimates the phase errors of the associated phase shifter, and uses these errors to predistort the \ps configuration in a closed-loop manner.
Unfortunately, this method implies reducing the system \ac{dr} and angular resolution during the calibration frames, which might not be acceptable in the automotive context.
For this reason, a first multi-\ac{tx} calibration method was proposed in \cite{jeannin2022particular}, in which multiple \ac{tx}-channels could be enabled simultaneously during a \ps calibration frame, maintaining the angular resolution and \ac{dr} of the system at the cost of requiring specific \ac{ddm} codes during the calibration frame.
However, the specific \ac{ddm} codes proposed for the \ps calibration frame present a serious scalability problem as it will be demonstrated in this paper.
In \cite{jeannin2022particular}, the number of \ac{tx}-channels that can be realistically enabled during a calibration frame is limited to four.
Many automotive radar systems employing \ac{ddm} are now using more than four \ac{tx}-channels such that the proposed calibration codes would impact the angular resolution and \ac{dr} during the calibration frame.

In this work, we propose an improved \ps \ac{ddm} calibration code family that can scale realistically to any number of multiplexed \ac{tx}-channels.
This code family is inspired by our previous research on \ac{oddm} codes \cite{jeannin2023doppler} offering the possibility to design \ac{ddm} codes with higher inter-\ac{tx} spurs' isolation.



\todo{preview}


\section{Background and Related Work}

	\subsection{\ac{ddm} model with a non-ideal \ps}
		\begin{figure}[t]
	        \centerline{\includegraphics[width=0.22\textwidth]{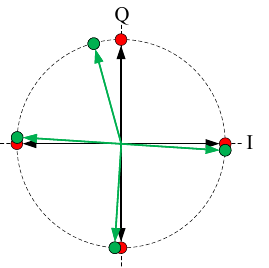}}
	        \begin{tikzpicture}[overlay]
	            \node[black, anchor=north west] at (5.8,2.2) {$\gls{constelIdx}{}_{\gls{txIdx}} = 0$};
	            \node[black, anchor=south east] at (4.1,4) {$\gls{constelIdx}{}_{\gls{txIdx}} = 1$};
	            \node[black, anchor=south east] at (2.7,2.3) {$\gls{constelIdx}{}_{\gls{txIdx}} = 2$};
	            \node[black, anchor=north east] at (4.2,0.75) {$\gls{constelIdx}{}_{\gls{txIdx}} = 3$};
	        \end{tikzpicture}
	        \caption{Illustration of an ideal \ac{qpsk} constellation (in red) and the associated erroneous constellation (in green). Note that the mean phase constellation error is zero ($\meanOf{\phaseOf{\gls{phaseErrVect}_{\gls{txIdx}}}} = 0$).}
	        \label{fig:qpsk erroneous constellation}
	    \end{figure}

		We define the \ac{tx} \ac{ddm} phase code sequence $\pc{\gls{txIdx}} [\gls{rampIdx}]$ as a function of the \ac{tx} constellation $\gls{constel}{}_{\gls{txIdx}} [ \gls{constelIdx}{}_{\gls{txIdx}} ]$ using the modulo operation following
		\begin{equation}\label{eq:ddm def}
			\pc{\gls{txIdx}}[\gls{rampIdx}] 
			= \gls{constel}{}_{\gls{txIdx}}[\gls{rampIdx} \bmod{\gls{pskNb}_{\gls{txIdx}}}]
			\text{,}
		\end{equation}
		where $\gls{txIdx}$ is the \ac{tx} index, $\gls{rampIdx} = 0, \hdots, \gls{rampNb}-1$ is the ramp number, and $\gls{constelIdx}{}_{\gls{txIdx}} = 0, \hdots, \Nk-1$ is the index of the constellation points (see Fig.~\ref{fig:qpsk erroneous constellation}).
		$\Nk$ is the \ac{psk} order (i.e., the number of points in the constellation).
	    The ideal \ac{ddm} constellation is defined as a function of the \ac{ddm} linear phase shift following
	    \begin{equation}\label{eq: ddm constellation}
	        \gls{constel}{}_{\gls{txIdx}} [ \gls{constelIdx}{}_{\gls{txIdx}} ] = \exp{\j \gls{constelIdx}{}_{\gls{txIdx}} \gls{ddmstep}_{\gls{txIdx}}}
	        \text{,}
	    \end{equation}
	    where $\j$ is the imaginary unit.
	    The \ac{ddm} linear phase shift $\gls{ddmstep}_{\gls{txIdx}}$ is defined as 
		\begin{equation}\label{eq: DDM phase step}
	        \gls{ddmstep}_{\gls{txIdx}} 
			= 2 \pi \frac{\gls{DDMbinShift}_{\gls{txIdx}} } {\gls{rampNb} }
	        = 2 \pi \frac{\gls{pskFactor}_{\gls{txIdx}} } { \gls{pskNb}_{\gls{txIdx}} }
	        \text{,}
	    \end{equation}
	    where $\gls{DDMbinShift}_{\gls{txIdx}} \in [0, \gls{rampNb} [$ is the resulting Doppler shift in bins and where $\gls{pskFactor}_{\gls{txIdx}} = 0,\hdots,\Nk-1$ is a factor that enables switching between the distinct \ac{ddm} shifts that can be achieved with a given \ac{psk} order.

	    We define the erroneous constellation as
	    \begin{equation}\label{eq:constellation example QPSK 2}
	        \err{\gls{constel}}{}_{\gls{txIdx}} [ \gls{constelIdx}{}_{\gls{txIdx}} ] =
	                \gls{constel}{}_{\gls{txIdx}} [ \gls{constelIdx}_{\gls{txIdx}} ]
	                \gls{phaseErr}_{\gls{txIdx}}[\gls{constelIdx}{}_{\gls{txIdx}}]
	        \text{,}
	    \end{equation}
	    using the complex error $\gls{phaseErr}_{\gls{txIdx}}[\gls{constelIdx}{}_{\gls{txIdx}}]$ associated with each point $\gls{constelIdx}{}_{\gls{txIdx}}$ of the constellation.
	    In general, we assume the mean phase error is zero $\meanOf{\phaseOf{\gls{phaseErrVect}_{\gls{txIdx}}}} = 0$ such that the \ps error does not inject any \ac{tx} channel imbalance \cite{jeannin2022modeling}.
	    Any non-zero mean phase error would translate into a channel phase imbalance which can be addressed by another calibration mechanism.
	    With this assumption, we avoid any interference of the proposed \ps online calibration method with any other online channel imbalance calibration mechanism \cite{achatz2022iterative}.

	    For an ideal constellation, the Doppler spectrum of the \ac{ddm} sequence is given by
	    \begin{equation}\label{eq:Dirichlet}
	        \spectr{\gls{pc}}_{\gls{txIdx}} [\gls{dopplerIdx}]
	            = \gls{rampNb}
	            \overbrace{\exp{ \j 2 \pi \gls{DDMbinShift}_{\gls{txIdx}}  \frac{(\gls{rampNb}-1)}{(2 \gls{rampNb})}}}^{\text{(3.)}}
	              \overbrace{\exp{-\j 2 \pi \gls{dopplerIdx}                 \frac{(\gls{rampNb}-1)}{(2 \gls{rampNb})}}}^{\text{(2.)}}
	              \overbrace{ \frac{\sin( 2 \pi (\gls{DDMbinShift}_{\gls{txIdx}} - \gls{dopplerIdx}) / 2)}
	                               { \sin(2 \pi (\gls{DDMbinShift}_{\gls{txIdx}} - \gls{dopplerIdx}) / (2 \gls{rampNb}) ) } 
	                        }^{\text{(1.)}}
	        \text{,}
	    \end{equation}
	    obtained using the Dirichlet kernel\footnote{The overbraced numbering will be used later to support the discussion.} \cite{jeannin2023doppler}.
	    Usually, the \ac{ddm} Doppler shift $\gls{DDMbinShift}_{\gls{txIdx}}$ is selected as a natural number such that the following simplification of \eqref{eq:Dirichlet} applies:
	    \begin{equation}\label{eq:DDM code spectrum 4}
	        \spectr{\gls{pc}}_{\gls{txIdx}} [\gls{dopplerIdx}] 
	            =   \begin{cases*}
	            \gls{rampNb} & if $ \gls{DDMbinShift}_{\gls{txIdx}} = \gls{dopplerIdx} $  \\
	            0 & if $ \gls{DDMbinShift}_{\gls{txIdx}} \neq \gls{dopplerIdx} $
	                \end{cases*}
	        \text{, for $\gls{DDMbinShift}_{\gls{txIdx}} \in (0,\hdots,\gls{rampNb}-1)$.}
	    \end{equation}
	    Using the simplification in \eqref{eq:DDM code spectrum 4}, we note the resulting \ac{ddm}-modulated \ac{rd} radar cube as
	    \begin{equation} \label{eq:SRD DDM final}
	        \gls{s_RD}{}_{\gls{modulated}} [\gls{rangeIdx},\gls{dopplerIdx},\gls{rxIdx}] = 
	            \sum_{\gls{txIdx} = 0}^{\gls{txNb}-1}
	                \gls{s_RD}_{,\gls{txIdx}} 
	                    [\gls{rangeIdx},(\gls{dopplerIdx}-\gls{DDMbinShift}_{\gls{txIdx}}) \bmod{\gls{rampNb}},\gls{rxIdx}]
	            \text{, for $\gls{DDMbinShift}_{\gls{txIdx}} \in \natural$,}
	    \end{equation}
	    where $\gls{rangeIdx} = 0,\hdots,\gls{ff1L}/2-1$ is the range bin index, $\gls{ff1L}$ is the \ac{rfft} length, $\gls{dopplerIdx} = 0,\hdots,\gls{rampNb}-1$ is the Doppler bin index, and $\gls{rxIdx} = 0,\hdots,\gls{rxNb}-1$ is the \ac{rx} index with $\gls{rxNb}$ the number of \ac{rx}-channels.
	    $\gls{s_RD}_{,\gls{txIdx}}$ is the single-\ac{tx} unmodulated radar cube containing the superposed signals of $\gls{targetNb}$ targets.
	    Eq. \eqref{eq:SRD DDM final} yields the classical \ac{ddm} \ac{rdm} with a single target duplicated $\gls{txNb}$ times over the Doppler spectrum with the \ac{ddm} Doppler shift in bins $\gls{DDMbinShift}_{\gls{txIdx}}$.
	    Each visible peak on the \ac{rdm} contains the complex channel information needed to assemble the channel vector exploited by the \ac{doa} estimator.

    \subsection{Exploiting the \ac{ddm} spurs' complex information}

	    \begin{figure}[t]
			\centerline{\includegraphics[]{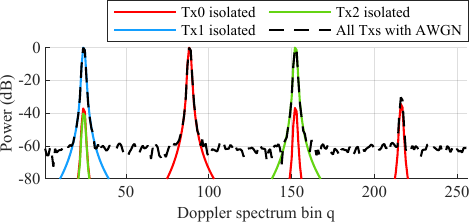}}
			\caption{Doppler cut from the \ac{rdm} of a given \ac{rx} at the range of a detected target for a 3-\ac{tx} \ac{ddm}-modulated system with $\gls{pskNb}_{(0,1,2)} = (4,4,4)$ and $\gls{pskFactor}_{(0,1,2)} = (1,0,2)$. The extracted signal is depicted in black while the isolated and noiseless influence of each transmitter $0,1,2$ is depicted in red, blue, and green respectively \cite{jeannin2022particular}.}
			\label{fig:multiTx DDM detailed}
		\end{figure}

	    \pgf{spurs location}
	    Let $\gls{targetIdx} = 0,\hdots,\gls{targetNb}-1$ be the index of a detected target and $\gls{dopplerIdx}_{\gls{targetIdx}}$ the associated Doppler bin index where the target was detected.
	    As per \cite{jeannin2022modeling}, for an erroneous constellation, each \ac{tx}-specific peak creates in turn spurs located at specific positions given by
	    \begin{equation}\label{eq:peak and spur group def}
	        \gls{psg}_{\gls{targetIdx}, \gls{txIdx}} [\gls{psgIdx}_{\gls{txIdx}}] 
	            = \left( 
	                \gls{dopplerIdx}_{\gls{targetIdx}} + 
	                \gls{psgIdx}_{\gls{txIdx}} \frac{\gls{pskFactor}_{\gls{txIdx}}}{\gls{pskNb}_{\gls{txIdx}}} \gls{rampNb}
	            \right) \bmod{\gls{rampNb}}
	        \text{,}
	    \end{equation}
	    where $\gls{psgIdx}_{\gls{txIdx}} = 0,\hdots,\Nk-1$ is the \psg index.
	    For instance, in Fig.~\ref{fig:multiTx DDM detailed}, for \ac{tx}$0$ (in red), the Doppler indices of the \psg are
	    \begin{equation*}
	    	\gls{psg}_{\gls{targetIdx}, \gls{txIdx}} [\gls{psgIdx}_{\gls{txIdx}} = (0,1,2,3)] = (28 , 92, 156, 220)
	    	\text{.}
	    \end{equation*}
	    Note that $\gls{psg}_{\gls{targetIdx}, \gls{txIdx}} [0]$ always points at the true target's velocity.

	    \pgf{DDM single tx estimation}

	    The \psg associated with a given \ac{tx}$\gls{txIdx}$ and a detected target $\gls{targetIdx}$ can be extracted from the \ac{rdm} and stored in a vector of length $\Nk$ whose $\gls{psgIdx}_{\gls{txIdx}}$th element is given by
	    \begin{equation}\label{eq:psg vector notation}
			\left[ \gls{psgVect}_{\gls{targetIdx},\gls{txIdx},\gls{rxIdx}} \right]
			_{\gls{psgIdx}_{\gls{txIdx}}}
			=
			\gls{s_RD}_{\gls{modulated}} [\gls{rangeIdx}_{\gls{targetIdx}},\gls{psg}_{\gls{targetIdx}, \gls{txIdx} } [\gls{psgIdx}_{\gls{txIdx}}],\gls{rxIdx}]
			\text{.}
		\end{equation}

		In \cite{jeannin2022particular}, it was shown that the \ps error of \ac{tx}$i$ can be estimated if the condition
		\begin{equation}\label{eq:multi Tx requirement}
			\frac{\gls{pskNb}_{i}}{\prod_{\substack{\gls{txIdx} = 0 \\ \gls{txIdx} \neq i}}^{\gls{txNb}-1}\Nk} 
			> 2 
			\text{,}
		\end{equation}
		is fulfilled, assuming that \eqref{eq:DDM code spectrum 4} is satisfied.
		This condition can be fulfilled in many cases by increasing $\gls{pskNb}_i$, which implies switching \ac{tx}$i$ to a higher \ac{psk} order.
		In the context of \ps calibration, the increased \ac{psk} order constellation should contain the original constellation, such that calibrating one would calibrate the other.
		As an example, consider \ac{tx}$0$ in the scenario in Fig.~\ref{fig:multiTx DDM detailed} (red).
		By switching the \ac{psk} order of \ac{tx}$0$ to 8-\ac{psk} ($\gls{pskNb}_0 = 8$), \eqref{eq:multi Tx requirement} is fulfilled ($8/(1 \cdot 2) > 2$) and the 8-\ac{psk} constellation (and the included \ac{qpsk} constellation) can be calibrated.
		This idea was successfully demonstrated in \cite{jeannin2022particular} in an automotive scenario for three \ac{tx}-channels.
	    However, in this paper, we are looking at the scalability of this idea.
	    
	    Taking the same assumptions as previously, and assuming $\gls{rampNb} \in \gls{po2}$, the set of power-of-two numbers, Tab.~\ref{tab:multi Tx idea scalability} gives the minimum required \ac{psk} order needed to satisfy \eqref{eq:multi Tx requirement}.
	    As it can be seen, the \ac{psk} order requirement is scaling exponentially leading to unrealistic requirements beyond four \ac{tx}-channels.


		\begin{table}[h]
	    \centering
	    \caption{Minimum valid \ac{psk} order for a given number of \ac{tx}-channels to satisfy \eqref{eq:multi Tx requirement}.}
	    \label{tab:multi Tx idea scalability}
	    \begin{tabular}{|c|c|l|}
	    \hline
	        \# \acp{tx} $\gls{txNb}$ 	& \makecell{Min valid \\ \ac{psk} order} & \multicolumn{1}{|c|}{Example} \\ \hline
	        2 			& 4 		& $\gls{pskNb}_{(0,1,\hdots)} = (4,1   )$\\ \hline
	        3 			& 8 		& $\gls{pskNb}_{(0,1,\hdots)} = (8,2,1 )$\\ \hline
	        4 			& 32 		& $\gls{pskNb}_{(0,1,\hdots)} = (32,4,2,1 	)$\\ \hline
	        5 			& 256 		& $\gls{pskNb}_{(0,1,\hdots)} = (256,4,4,2,1 	)$\\ \hline
	        6 			& 1024 		& $\gls{pskNb}_{(0,1,\hdots)} = (1024,8,4,4,2,1 	)$\\ \hline
	    \end{tabular}
	    \end{table}


\section{Methodology}
	\subsection{ODDM}
		The exponential scaling of the \ac{psk} order originates from assumption \eqref{eq:DDM code spectrum 4}.
		In \cite{jeannin2023doppler}, we analyzed the implication of using $\gls{DDMbinShift}_{\gls{txIdx}} \notin \natural$.
		This implies dealing with \eqref{eq:Dirichlet} and compensating the three overbraced terms:
	    \begin{enumerate}
	        \item The magnitude distortion related to the windowing effect.
	        \item The linear phase distortion along the Doppler spectrum related to the \ac{dft} definition.
	        \item The \ac{tx} channel phase distortion related to the starting phase of the \ac{ddm} sequence (usually $0\deg$).
	    \end{enumerate}
	    The \ac{ddm} codes using $\gls{DDMbinShift}_{\gls{txIdx}} \notin \natural$ are called \ac{oddm}.

	\subsection{Improving the \psg isolation}

		\begin{figure}[t]
		     \centering
		     \begin{subfigure}{0.48\textwidth}
		         \centering
		         \includegraphics[]{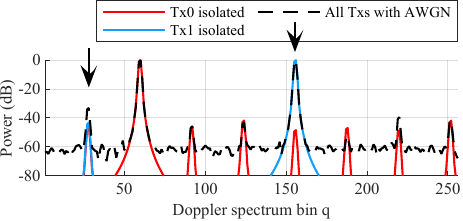}
		         \caption{}
		         \label{fig:bpsk}
		     \end{subfigure}
		     \hfill
		     \begin{subfigure}{0.48\textwidth}
		         \centering
		         \includegraphics[]{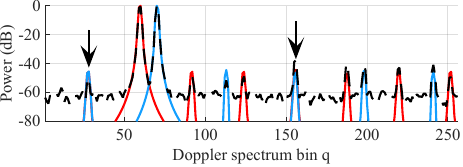}
		         \caption{}
		         \label{fig:6psk}
		     \end{subfigure}
		    \caption{Doppler cut from the \ac{rdm} of a given \ac{rx} at the range of a detected target for a 2-\ac{tx} \ac{ddm}-modulated system (a) $\gls{pskNb}_{(0,1)} = (8,2)$ and $\gls{pskFactor}_{(0,1)} = (1,1)$ and (b) $\gls{pskNb}_{(0,1)} = (8,6)$ and $\gls{pskFactor}_{(0,1)} = (1,1)$. In both cases, despite the different \ac{ddm} codes, only two spurs overlap.}
		    \label{fig:bpsk 6psk}
		\end{figure}

		Let us consider a 2-\ac{tx} \ac{ddm}-modulated system with $\gls{pskNb}_{(0,1)} = (8,2)$.
		We select $\gls{rampNb} = 256$ such that $\gls{DDMbinShift}_{(0,1)} = (32,128) \in \natural$.
		This use case is depicted in Fig.~\ref{fig:bpsk} for a single target where the two arrows designate the spurs overlap from \ac{tx}$0$ and \ac{tx}$1$.
		As per \eqref{eq:multi Tx requirement}, the \psg of \ac{tx}$0$ contains sufficient information for the \ps error estimation.

		Let us now consider $\gls{pskNb}_{(0,1)} = (8,6)$.
		With the same parameters as previously defined, $\gls{DDMbinShift}_{1} = 46.\bar{6} \notin \natural$.
		This use case is depicted in Fig.~\ref{fig:6psk} for a single target where the two arrows designate the spurs overlap from \ac{tx}$0$ and \ac{tx}$1$.
		Interestingly, we note that the spur overlap is the same as in the previous case.
		Consequently, the \psg of \ac{tx}$0$ should contain sufficient information for the \ps error estimation.

		The link between the first and the second case can be formalized using the \ac{gcd}:
		\begin{equation}\label{eq:gcd}
			\gcd(8,2) = \gcd(8,6) = 2
			\text{,}
		\end{equation}
		such that, by dropping the assumption in \eqref{eq:DDM code spectrum 4}, \eqref{eq:multi Tx requirement} can be relaxed to
		\begin{equation}\label{eq:multi Tx coprime requirement}
			\frac{\gls{pskNb}_{i}}
				{\prod_{\substack{\gls{txIdx} = 0 \\ \gls{txIdx} \neq i}}^{\gls{txNb}-1} 
					\text{gcd} ( \gls{pskNb}_{i} , \Nk ) } 
			> 2 
			\text{.}
		\end{equation}

		Taking the same assumptions as previously, and assuming $\gls{rampNb} \in \gls{po2}$, Tab.~\ref{tab:multi Tx idea scalability} gives the minimum required \ac{psk} order needed to satisfy \eqref{eq:multi Tx coprime requirement}.
	    As it can be seen, the \ac{psk} order requirement is scaling linearly leading to realistic \ac{psk} requirements beyond four \ac{tx}-channels.

	    \begin{table}[h]
	    \centering
	    \caption{Minimum valid \ac{psk} order for a given number of \ac{tx}-channels to satisfy \eqref{eq:multi Tx coprime requirement}.}
	    \label{tab:multi Tx idea coprime scalability}
	    \begin{tabular}{|c|c|l|}
	    \hline
	        \# \acp{tx} $\gls{txNb}$ 	& \makecell{Min. valid \\ \ac{psk} order} & \multicolumn{1}{|c|}{Example}\\ \hline
	        2 			& 3 		& $\gls{pskNb}_{(0,1,\hdots)} = (3,1)$			\\ \hline
	        3 			& 3 		& $\gls{pskNb}_{(0,1,\hdots)} = (3,2,1)$		\\ \hline
	        4 			& 4 		& $\gls{pskNb}_{(0,1,\hdots)} = (4,3,3,1)$		\\ \hline
	        5 			& 5 		& $\gls{pskNb}_{(0,1,\hdots)} = (5,4,3,1)$		\\ \hline
	        6 			& 6 		& $\gls{pskNb}_{(0,1,\hdots)} = (6,5,5,5,2,1)$	\\ \hline
	    \end{tabular}
	    \end{table}

	\subsection{Proposed calibration sequence}

		In \cite{jeannin2022particular}, we suggested sequentially switching the \ac{psk} order of the \ac{tx} of interest to a higher order to create sufficient isolation from the other \ac{tx}-channels.
		In this paper, we suggest keeping the \ac{tx} of interest at its operational \ac{psk} order and switching the other \ac{tx}-channels \ac{psk} orders to satisfy \eqref{eq:multi Tx coprime requirement}.
		This method is much more convenient and efficient.

		\begin{figure}[t]
			\centerline{\includegraphics[]{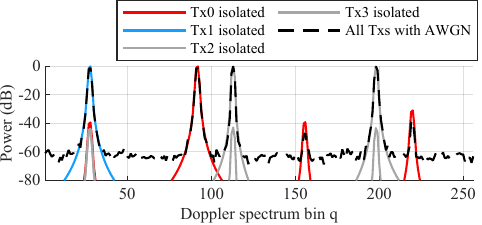}}
			\caption{Doppler cut from the \ac{rdm} of a given \ac{rx} at the range of a detected target for a 4-\ac{tx} \ac{ddm}-modulated system with $\gls{pskNb}_{(0,1,2,3)} = (4,4,3,3)$ and $\gls{pskFactor}_{(0,1,2,3)} = (1,0,1,2)$. The extracted signal is depicted in black while the isolated and noiseless influence of each transmitter $0,1,2,4$ is depicted in red, blue, gray, and gray respectively.}
			\label{fig:multiTx ODDM detailed}
		\end{figure}

		To demonstrate this new method, let us consider a 4-\ac{tx} \ac{ddm}-modulated system with $\gls{pskNb}_{(0,1,2,3)} = (1,8,4,2)$ and $\gls{pskFactor}_{(0,1,2,3)} = (1,1,1,1)$ leading to $\gls{DDMbinShift}_{(0,1,2,3)} = (0,32,64,96)$ for $\gls{rampNb} = 256$.
		Assuming \ac{tx}$2$ as the \ac{tx} of interest
		\begin{equation}\label{eq:multi Tx coprime requirement 1}
			\frac{\gls{pskNb}_{i}}
				{\prod_{\substack{\gls{txIdx} = 0 \\ \gls{txIdx} \neq i}}^{\gls{txNb}-1} 
					\text{gcd} ( \gls{pskNb}_{i} , \Nk ) } 
			= 0.5 \leq 2
			\text{,}
		\end{equation}
		which indicates that the \psg cannot be used for \ac{tx}$2$ \ps error estimation.

		To calibrate \ac{tx}$2$, we can switch $\gls{pskNb}_{(0,1,2,3)} = (1,3,4,3)$ and $\gls{pskFactor}_{(0,1,2,3)} = (1,1,1,2)$ leading to $\gls{DDMbinShift}_{(0,1,2,3)} = (0,85.\bar{3},64,170.\bar{6})$ for $\gls{rampNb} = 256$.
		This use case is depicted in Fig.~\ref{fig:multiTx ODDM detailed} for a single isolated target.
		With this calibration \ac{ddm} code, assuming \ac{tx}$2$ as the \ac{tx} of interest 
		\begin{equation}\label{eq:multi Tx coprime requirement 2}
			\frac{\gls{pskNb}_{i}}
				{\prod_{\substack{\gls{txIdx} = 0 \\ \gls{txIdx} \neq i}}^{\gls{txNb}-1} 
					\text{gcd} ( \gls{pskNb}_{i} , \Nk ) } 
			= 4 > 2
			\text{,}
		\end{equation}
		which indicates that the \psg can be used for \ac{tx}$2$ \ps error estimation.

		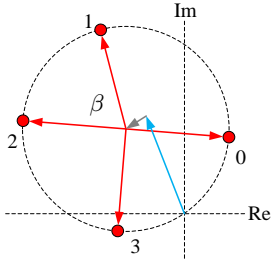
\begin{figure}[t]
			\centerline{\includegraphics[width=0.22\textwidth]{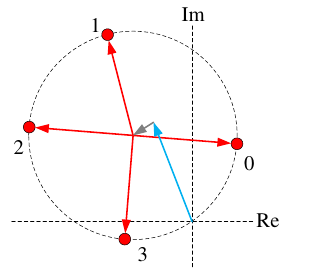}}
			\begin{tikzpicture}[overlay, every node/.style={font=\small}]
	            \node[anchor= north east] at (4,2.75) {$\gls{barycenter}$};
	        \end{tikzpicture}
			\caption{Illustration of the super-constellation obtained by processing $\gls{mF}^{-1}_{4} \gls{psgVect}_{\gls{targetIdx},2,\gls{rxIdx}}$ extracted from the Doppler cut depicted in Fig.~\ref{fig:multiTx ODDM detailed}. The influence of each \ac{tx} in the construction of the super-constellation is depicted using the same color coding as in Fig.~\ref{fig:multiTx ODDM detailed}.}
			\label{fig:multiTx ODDM constel}
		\end{figure}

		Indeed, following the method proposed in \cite{jeannin2022particular}, the so-called super-constellation can be built from the \psg of \ac{tx}$2$ following $\gls{mF}^{-1}_{\gls{pskNb}_2} \gls{psgVect}_{\gls{targetIdx},2,\gls{rxIdx}}$, where $\gls{mF}^{-1}_{\gls{pskNb}_2}$ is the $\gls{pskNb}_2$-point \ac{idft} matrix.
		The resulting super-constellation is illustrated in Fig.~\ref{fig:multiTx ODDM constel} with the complex contribution of the other \ac{tx}-channels depicted with the same color coding as in Fig.~\ref{fig:multiTx ODDM detailed}.
		The erroneous \ac{qpsk} constellation of \ac{tx}$2$ appears clearly in this figure and can be estimated following
		\begin{equation}\label{eq:multi tx example 2tx estimation}
			\est{\gls{phaseErrVect}}_{2}
			= 
			\gls{constelVect}_{2}^{\ast}
			\had
			\left(
			\gls{mF}^{-1}_{4}
			\gls{psgVect}_{\gls{targetIdx},2,\gls{rxIdx}}
			- \bm{1} \est{\gls{barycenter}}
			\right)
			\frac{1}{\est{\gls{alpha_RD}}_{ \gls{targetIdx},2,\gls{rxIdx}}}
			\text{.}
		\end{equation}
		where $\gls{constelVect}_{2}$ is a vector of length $\gls{pskNb}_2$ containing the \ac{tx}$2$ \ps constellation points, and where $\est{\gls{phaseErrVect}}_{2}$ is a vector containing the $\gls{pskNb}_2$ estimates of $\gls{phaseErr}_{2}[\gls{constelIdx}_{2}]$.
		$\est{\gls{barycenter}}$ is the estimated center of the red \ac{tx}$2$ constellation multiplied by the unit vector $\bm{1}$, $\gls{alpha_RD}_{ \gls{targetIdx},2,\gls{rxIdx}}$ is the normalization factor that can be estimated, and $\left[ \gls{psgVect}_{\gls{targetIdx},2,\gls{rxIdx}} \right]_1$ is the complex value of the target peak associated with \ac{tx}$2$ \cite{jeannin2022particular}.
		$\had$ is the Hadamard product operator.



\section{Simulation and Results}

	\begin{figure}[t]
		\centerline{\includegraphics[]{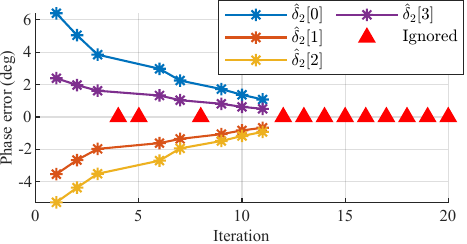}}
		\caption{Evolution of the estimated phase errors of $\est{\gls{phaseErrVect}}_{2}$ from the \ac{tx}$2$ \ac{qpsk} constellation during a 20-frame closed-loop calibration simulation of the multi-\ac{tx} setup depicted in Fig.~\ref{fig:multiTx ODDM detailed} using the same simulation setup as \cite{jeannin2022modeling}. Each colored line indicates the estimated phase error of a given point of the \ac{tx}$2$ \ac{qpsk} constellation. A red triangle indicates that no spurs were detected in the \ac{rdm}.}
		\label{fig:simulation}
	\end{figure}

	Using the \ac{ddm} simulation framework presented in \cite{jeannin2022modeling}, we simulated the closed-loop predistortion calibration of \ac{tx}$2$ in a 4-\ac{tx} \ac{ddm}-modulated system following the configuration illustrated in Fig.~\ref{fig:multiTx ODDM detailed}.
	The result is given in Fig.~\ref{fig:simulation} showing a convergence of the individual phase error components of $\est{\gls{phaseErrVect}}_{2}$ towards zero until the \ps error is sufficiently small such that no spurs are visible on the \ac{rdm}, validating the correctness of the \ps error estimation.



\section{Conclusion}
	In this paper, we presented a new online \ps calibration \ac{ddm} sequence that allows scaling the system to any number of simultaneously activated \ac{tx}-channels realistically during the calibration period.
	The angular resolution and \ac{dr} of the system are maintained during the online \ps calibration independently from the number of \ac{tx}-channels in the system.
	We achieved this by exploiting \ac{oddm} as a way to reduce the inter-\ac{tx} multiplexing leakage.




\bibliographystyle{IEEEtran_isp}

\bibliography{literature}

\end{document}

%% file: EuMW_modify_IEEEtran_18b_CTAN_v5.tex
\makeatletter

\def\@maketitle{\newpage
\bgroup\par\addvspace{0.5\baselineskip}\centering%
\ifCLASSOPTIONtechnote
   {\bfseries\large\@IEEEcompsoconly{\sffamily}\@title\par}\vskip 1.3em{\lineskip .5em\@IEEEcompsoconly{\sffamily}\@author
   \@IEEEspecialpapernotice\par{\@IEEEcompsoconly{\vskip 1.5em\relax
   \@IEEEtitleabstractindextextbox{\@IEEEtitleabstractindextext}\par
   \hfill\@IEEEcompsocdiamondline\hfill\hbox{}\par}}}\relax
\else
   \vskip0.2em{\EuMWtitlesize\ifCLASSOPTIONtransmag\bfseries\LARGE\fi\@IEEEcompsoconly{\sffamily}\@IEEEcompsocconfonly{\normalfont\normalsize\vskip 2\@IEEEnormalsizeunitybaselineskip
   \bfseries\Large}\@title\par}\vskip1.0em\par
   \ifCLASSOPTIONconference%
      {\@IEEEspecialpapernotice\mbox{}\vskip\@IEEEauthorblockconfadjspace%
       \mbox{}\hfill\begin{@IEEEauthorhalign}\@author\end{@IEEEauthorhalign}\hfill\mbox{}\par}\relax
   \else
      \ifCLASSOPTIONpeerreviewca
         {\@IEEEcompsoconly{\sffamily}\@IEEEspecialpapernotice\mbox{}\vskip\@IEEEauthorblockconfadjspace%
          \mbox{}\hfill\begin{@IEEEauthorhalign}\@author\end{@IEEEauthorhalign}\hfill\mbox{}\par
          {\@IEEEcompsoconly{\vskip 1.5em\relax
           \@IEEEtitleabstractindextextbox{\@IEEEtitleabstractindextext}\par\hfill
           \@IEEEcompsocdiamondline\hfill\hbox{}\par}}}\relax
      \else
         \ifCLASSOPTIONtransmag
           {\@IEEEspecialpapernotice\mbox{}\vskip\@IEEEauthorblockconfadjspace%
            \mbox{}\hfill\begin{@IEEEauthorhalign}\@author\end{@IEEEauthorhalign}\hfill\mbox{}\par
           {\vspace{0.5\baselineskip}\relax\@IEEEtitleabstractindextextbox{\@IEEEtitleabstractindextext}\vspace{-1\baselineskip}\par}}\relax
         \else
           {\lineskip.5em\@IEEEcompsoconly{\sffamily}\sublargesize\@author\@IEEEspecialpapernotice\par
           {\@IEEEcompsoconly{\vskip 1.5em\relax
            \@IEEEtitleabstractindextextbox{\@IEEEtitleabstractindextext}\par\hfill
            \@IEEEcompsocdiamondline\hfill\hbox{}\par}}}\relax
         \fi
      \fi
   \fi
\fi\par\addvspace{0.0\baselineskip}\egroup}

\def\EuMWtitlesize{\@setfontsize{\EuMWtitlesize}{24}{24pt}}
\def\EuMWauthorsize{\@setfontsize{\EuMWauthorsize}{11}{11pt}}
\def\EuMWaffilsize{\@setfontsize{\EuMWaffilsize}{10}{10pt}}
\def\EuMWcaptionsize{\@setfontsize{\EuMWcaptionsize}{9}{10pt}}
\def\EuMWbibsize{\@setfontsize{\EuMWbibsize}{8}{10pt}}

\def\@IEEEauthorblockNstyle{\EuMWauthorsize\@IEEEcompsocnotconfonly{\sffamily}\@IEEEcompsocconfonly{\large}}
\def\@IEEEauthorblockAstyle{\EuMWaffilsize\@IEEEcompsocnotconfonly{\sffamily}\@IEEEcompsocconfonly{\itshape}\@IEEEcompsocconfonly{\large}}
\def\@IEEEauthordefaulttextstyle{\EuMWauthorsize\@IEEEcompsocnotconfonly{\sffamily}\sublargesize}

\def\thebibliography#1{\section*{\refname}%
    \addcontentsline{toc}{section}{\refname}%
    \EuMWbibsize\@IEEEcompsocconfonly{\small}\vskip 0.3\baselineskip plus 0.1\baselineskip minus 0.1\baselineskip
    \list{\@biblabel{\@arabic\c@enumiv}}%
    {\settowidth\labelwidth{\@biblabel{#1}}%
    \leftmargin\labelwidth
    \advance\leftmargin\labelsep\relax
    \itemsep \IEEEbibitemsep\relax
    \usecounter{enumiv}%
    \let\p@enumiv\@empty
    \renewcommand\theenumiv{\@arabic\c@enumiv}}%
    \let\@IEEElatexbibitem\bibitem%
    \def\bibitem{\@IEEEbibitemprefix\@IEEElatexbibitem}%
\def\newblock{\hskip .11em plus .33em minus .07em}%
\ifCLASSOPTIONtechnote\sloppy\clubpenalty4000\widowpenalty4000\interlinepenalty100%
\else\sloppy\clubpenalty4000\widowpenalty4000\interlinepenalty500\fi%
    \sfcode`\.=1000\relax}

\long\def\@makecaption#1#2{%
\ifx\@captype\@IEEEtablestring%
\par\@IEEEtabletopskipstrut
\else
\@IEEEfigurecaptionsepspace
\fi
\setbox\@tempboxa\hbox{\normalfont\footnotesize {#1.}\nobreakspace\nobreakspace #2}%
\ifdim \wd\@tempboxa >\hsize%
\setbox\@tempboxa\hbox{\normalfont\footnotesize {#1.}\nobreakspace\nobreakspace}%
\parbox[t]{\hsize}{\normalfont\footnotesize\noindent\unhbox\@tempboxa#2}%
\else
\ifCLASSOPTIONconference \hbox to\hsize{\normalfont\footnotesize\hfil\box\@tempboxa\hfil}%
\else \hbox to\hsize{\normalfont\footnotesize\box\@tempboxa\hfil}%
\fi\fi
\ifx\@captype\@IEEEtablestring%
\@IEEEtablecaptionsepspace
\else
\fi}

\newlength\tablecaptiontotableskip
\newlength\figuretocaptionskip
\def\@IEEEfigurecaptionsepspace{\vskip\figuretocaptionskip\relax}%
\def\@IEEEtablecaptionsepspace{\vskip\tablecaptiontotableskip\relax}%

\def\abstract{\normalfont%
\@IEEEabskeysecsize\bfseries\textit{\abstractname}\,\bfseries\textit{---}\,%
\@IEEEgobbleleadPARNLSP}%

\def\IEEEkeywords{\normalfont%
\@IEEEabskeysecsize\bfseries\textit{\IEEEkeywordsname}\,\bfseries\textit{---}\,%
\@IEEEgobbleleadPARNLSP}%
\def\endIEEEkeywords{\relax\vspace{0.67ex}%
\par\if@twocolumn\else\endquotation\fi%
\normalsize\normalfont}%

\DeclareRobustCommand*{\EuMWauthorrefmark}[1]{\raisebox{0pt}[0pt][0pt]{\textsuperscript{#1}}}%
\def\@IEEEauthorblockNtopspace{0ex}
\def\@IEEEauthorblockAtopspace{1mm}
\def\tablename{Table}
\def\thetable{\arabic{table}}
\def\IEEEkeywordsname{Keywords}
\def\subsubsection{\@startsection{subsubsection}{3}{\z@}{1.5ex plus 1.5ex minus 0.5ex}%
{0.7ex plus .5ex minus 0ex}{\normalfont\normalsize\itshape}}%
\newlength{\CPheadmatchindent}%
\def\@seccntformat#1{\hbox to\CPheadmatchindent{\csname the#1dis\endcsname}\hskip 0.1em \relax}
\IEEEilabelindentA \parindent
\IEEEilabelindent \IEEEilabelindentA
\IEEEelabelindent \parindent
\IEEEdlabelindent \parindent
\IEEElabelindent \parindent
\makeatother

%% file: commands.tex
\DeclareFontFamily{U}{wncy}{}
\DeclareFontShape{U}{wncy}{m}{n}{<->wncyr10}{}
\DeclareSymbolFont{mcy}{U}{wncy}{m}{n}
\DeclareMathSymbol{\Sh}{\mathord}{mcy}{"58}

\newcommand{\ps}{phase shifter\xspace} %
\newcommand{\pss}{phase shifters\xspace} %
\newcommand{\psg}{peak-and-spur group\xspace} %
\newcommand{\Psg}{Peak-and-spur group\xspace} %

\ifrelease
	\newcommand{\todo}[1]{}
	\newcommand{\new}[1]{}
	\newcommand{\note}[1]{}
	\newcommand{\disk}[1]{}
	\newcommand{\copied}[1]{}
	\newcommand{\pgf}[1]{}

	\newcommand{\textttCom}[1]{}

\else
	\newcommand{\todo}[1]{{\color{red}TODO: #1}\\}
	\newcommand{\new}[1]{{\color{blue}#1}}
	\newcommand{\note}[1]{{\color{blue}NOTE: #1}\\}
	\newcommand{\disk}[1]{{\color{violet}#1}}
	\newcommand{\copied}[1]{{\color{gray}Copied from paper:\\#1}}
	\newcommand{\pgf}[1]{{\color{red}§ #1\\}}

	\newcommand{\textttCom}[1]{\texttt{#1}} 
	
\fi

\newcommand{\m}{\mathbf}
\newcommand{\mg}[1]{\bm{#1}} 
\newcommand{\err}[1]{\check{#1}}
\newcommand{\est}[1]{\widehat{#1}}
\newcommand{\corr}[1]{\ddot{#1}}
\renewcommand{\deg}{^\circ}

\newcommand{\pc}[1]{\gls{pc}_{#1}} 

\newcommand{\had}{\gls{prod}} 
\renewcommand{\j}{\gls{i}}

\renewcommand{\natural}{\gls{natural}}

\newcommand{\imOf}{\Im}
\newcommand{\realOf}{\Re}
\newcommand{\phaseOf}{\angle}

\newcommand{\meanOf}[1]{\overline{#1}}

\newcommand{\conj}[1]{{#1}^{\ast}}

\newcommand{\spectr}[1]{\widetilde{#1}} 
\renewcommand{\exp}[1]{\gls{exp}{}^{#1}} 

\newcommand{\nquad}[1]{\ifnum #1>0 \quad\nquad{\numexpr #1-1\relax}\fi}
\newcommand{\tdmIdx}{\gls{rampIdx}_{\gls{modulated}}}
\newcommand{\Nk}{\gls{pskNb}_{\gls{txIdx}}}

%% file: glossary.tex

\newglossaryentry{i}%
{%
  name={$j$},
  text={j},
  description={Imaginary unit \textttCom{(i)}},
}

\newglossaryentry{conv}%
{%
  name={$*$},
  text={*},
  description={Discrete convolution operator \textttCom{(conv)}},
}

\newglossaryentry{prod}%
{%
  name={$\odot$},
  text={\odot},
  description={Hadamard product operator \textttCom{(prod)}},
}

\newglossaryentry{mod}%
{%
  name={$\%$},
  text={\%},
  description={Modulo operator \textttCom{(mod)}},
}

\newglossaryentry{exp}%
{%
  name={$\text{e}$},
  text={\text{e}},
  description={Euler's number \textttCom{(exp)}},
}

\newglossaryentry{expm}%
{%
  name={$\textbf{e}$},
  text={\textbf{e}},
  description={Element-wise exponential \textttCom{(expm)}},
}

\newglossaryentry{conjugate}%
{%
  name={$\conj{x}$},
  text={\conj{x}},
  description={Conjugate of $x$ \textttCom{(conjugate)}},
}

\newglossaryentry{contx}%
{%
  name={$(x)$},
  text={(x)},
  description={Continuous variable \textttCom{(contx)}},
}

\newglossaryentry{est}%
{%
  name={$\est{x}$},
  text={\est{x}},
  description={Estimated parameter \textttCom{(est)}},
}

\newglossaryentry{err}%
{%
  name={$\err{x}$},
  text={\err{x}},
  description={Erroneous parameter \textttCom{(err)}},
}

\newglossaryentry{corr}%
{%
  name={$\corr{x}$},
  text={\corr{x}},
  description={Corrected parameter \textttCom{(corr)}},
}

\newglossaryentry{discx}%
{%
  name={$[x]$},
  text={[x]},
  description={Discrete variable \textttCom{(discx)}},
}

\newglossaryentry{x}%
{%
  name={$\m x$},
  text={\m x},
  description={Vector variable \textttCom{(x)}},
}

\newglossaryentry{x_transposed}%
{%
  name={$\m x ^{\m T}$},
  text={\m x ^{\m T}},
  description={Transposed vector \textttCom{(x\_transposed)}},
}

\newglossaryentry{X}%
{%
  name={$\m X$},
  text={\m X},
  description={Matrix variable \textttCom{(X)}},
}

\newglossaryentry{scalarFromX}%
{%
  name={$[\m X]_{x,y}$},
  text={[\m X]},
  description={Scalar extracted from $\m X$ at position $x,y$ \textttCom{(scalarFromX)}},
}

\newglossaryentry{F}%
{%
  name={$\mathcal{F}_{N}$},
  text={\mathcal{F}},
  description={$N$-point \acs{dft} operator \eqref{eq:DFT function}\textttCom{(F)}},
}

\newglossaryentry{rou}%
{%
  name={$\omega_{N}$},
  text={\omega},
  description={Primitive $N$th root of unity \eqref{eq:dft matrix def}\textttCom{(rou)}},
}

\newglossaryentry{spectr}%
{%
  name={$\spectr{\gls{x}}$},
  text={\spectr{\gls{x}}},
  description={\acs{dft} of vector $\gls{x}$ (Sec.~\ref{sec:intro:notation})\textttCom{(spectr)}},
}


\newglossaryentry{comp}%
{%
  name={$\mathbb{C}$},
  text={\mathbb{C}},
  description={The set of complex numbers \textttCom{(comp)}},
}

\newglossaryentry{real}%
{%
  name={$\mathbb{R}$},
  text={\mathbb{R}},
  description={The set of real numbers \textttCom{(real)}},
}

\newglossaryentry{natural}%
{%
  name={$\mathbb{N}$},
  text={\mathbb{N}},
  description={The set of natural numbers \textttCom{(natural)}},
}

\newglossaryentry{po2}%
{%
  name={$\mathbb{B}$},
  text={\mathbb{B}},
  description={The set of power-of-two numbers \eqref{eq:po2 def}\textttCom{(po2)}},
}

\newglossaryentry{muk}%
{%
  name={$\mathbb{M}_{\gls{constelIdx}{}_{\gls{txIdx}}}$},
  text={\mathbb{M}_{\gls{constelIdx}{}_{\gls{txIdx}}}},
  description={The sub-set of ramp indices (Sec.~\ref{subsec:autofocus:Error estimation})\textttCom{(po2)}},
}

\newglossaryentry{Muk}%
{%
  name={$\gls{rampNb}_{\gls{constelIdx}_{\gls{txIdx}}}$},
  text={\gls{rampNb}_{\gls{constelIdx}_{\gls{txIdx}}}},
  description={Subset size (Sec.~\ref{subsec:autofocus:Error estimation})\textttCom{(po2)}},
}

\newglossaryentry{imOf}%
{%
  name={$\imOf$},
  text={\imOf},
  description={Imaginary part \textttCom{(imOf)}},
}

\newglossaryentry{realOf}%
{%
  name={$\realOf$},
  text={\realOf},
  description={Real part \textttCom{(realOf)}},
}

\newglossaryentry{phaseOf}%
{%
  name={$\phaseOf$},
  text={\phaseOf},
  description={Phase of a complex number \textttCom{(phaseOf)}},
}

\newglossaryentry{absOf}%
{%
  name={$\lvert \cdot \rvert$},
  text={\lvert \cdot \rvert},
  description={Absolute value \textttCom{(absOf)}},
}

\newglossaryentry{floorOf}%
{%
  name={$\lfloor \cdot \rfloor$},
  text={\lfloor \cdot \rfloor},
  description={Floor value \textttCom{(floorOf)}},
}

\newglossaryentry{divides}%
{%
  name={$a \mid b$},
  text={\mid},
  description={$a$ divides $b$ \textttCom{(divides)}},
}

\newglossaryentry{ndivides}%
{%
  name={$a \nmid b$},
  text={\nmid},
  description={$a$ does not divide $b$ \textttCom{(ndivides)}},
}


\newglossaryentry{range}
{%
  name={$R_{\gls{targetIdx}} (t)$},
  text={R},
  description={Target's range evolution (m) (Sec.~\ref{eq:range velocity migration})\textttCom{(range)}},
}

\newglossaryentry{range0}
{%
  name={$R_{0,\gls{targetIdx}}$},
  text={R_{0}},
  description={Initial target's range (m) (Sec.~\ref{sec:analysis:target model})\textttCom{(range0)}},
}

\newglossaryentry{doppler}
{%
  name={$V_{\gls{targetIdx}}$},
  text={V},
  description={Target's radial velocity ($\text{m}\cdot\text{s}^{-1}$) (Sec.~\ref{sec:analysis:target model}) \textttCom{(doppler)}},
}

\newglossaryentry{rtd}
{%
  name={$\tau$},
  text={\tau},
  description={Round trip delay (s) \eqref{eq:round trip}\textttCom{(rtd)}},
}

\newglossaryentry{doa}%
{%
  name={$\theta_{\gls{targetIdx}}$},
  text={\theta},
  description={Direction of arrival angle (rad) (Sec.~\ref{sec:analysis:target model})\textttCom{(doa)}},
}

\newglossaryentry{vego}
{%
  name={$\gls{doppler}{}_{\text{ego}}$},
  text={\gls{doppler}{}_{\text{ego}}},
  description={Ego-velocity ($\text{m}\cdot\text{s}^{-1}$) \eqref{eq:doa cosine} \textttCom{(vego)}},
}

\newglossaryentry{mountAng}%
{%
  name={$\theta_{\text{ego}}$},
  text={\theta_{\text{ego}}},
  description={Mount angle (rad) \eqref{eq:doa cosine} \textttCom{(mountAng)}},
}


\newglossaryentry{targetIdx}%
{%
  name={$p$},
  text={p},
  description={Target index (Sec.~\ref{sec:analysis:target model})\textttCom{(targetIdx)}},
}

\newglossaryentry{targetNb}%
{%
  name={$P$},
  text={P},
  description={Target number (Sec.~\ref{sec:analysis:target model})\textttCom{(targetNb)}},
}

\newglossaryentry{sampleNb}%
{%
  name={$N$},
  text={N},
  description={Sample number (Sec.~\ref{subsec:analysis:sampling})\textttCom{(sampleNb)}},
}

\newglossaryentry{sampleIdx}%
{%
  name={$n$},
  text={n},
  description={Sample index (Sec.~\ref{subsec:analysis:sampling})\textttCom{(sampleIdx)}},
}

\newglossaryentry{rampNb}%
{%
  name={$M$},
  text={M},
  description={Ramp number (Sec.~\ref{sec:analysis:Resolving the velocity})\textttCom{(rampNb)}},
}

\newglossaryentry{rampIdx}%
{%
  name={$m$},
  text={m},
  description={Ramp index (Sec.~\ref{sec:analysis:Resolving the velocity})\textttCom{(rampIdx)}},
}

\newglossaryentry{ff1L}%
{%
  name={$N_{1}$},
  text={N_{1}},
  description={Range \acs{fft} length \eqref{eq:SR}\textttCom{(ff1L)}},
}

\newglossaryentry{rangeIdx}%
{%
  name={$r$},
  text={r},
  description={Range bin index \eqref{eq:SR}\textttCom{(rangeIdx)}},
}

\newglossaryentry{ff2L}%
{%
  name={$N_{2}$},
  text={N_{2}},
  description={Doppler \acs{fft} length \eqref{eq:SRD}\textttCom{(ff2L)}},
}

\newglossaryentry{dopplerIdx}%
{%
  name={$q$},
  text={q},
  description={Doppler bin index \eqref{eq:SRD}\textttCom{(dopplerIdx)}},
}

\newglossaryentry{angleIdx}%
{%
  name={$\rho$},
  text={\rho},
  description={Angle bin index \eqref{eq:S_RDA}\textttCom{(angleIdx)}},
}

\newglossaryentry{chNb}
{%
  name={$O$},
  text={O},
  description={Virtual channel number \eqref{eq:channel idx def}\textttCom{(chNb)}},
}

\newglossaryentry{chIdx}
{%
  name={$o$},
  text={o},
  description={Virtual channel index \eqref{eq:channel idx def}\textttCom{(chIdx)}},
}

\newglossaryentry{txNb}%
{%
  name={$K$},
  text={K},
  description={\acs{tx} channel number (Sec.~\ref{sec:analysis_mimo:advantage of multi tx an rx})\textttCom{(txNb)}},
}

\newglossaryentry{txIdx}%
{%
  name={$k$},
  text={k},
  description={\acs{tx} channel index (Sec.~\ref{subsec:analysis:Received RF signal})\textttCom{(txIdx)}},
}

\newglossaryentry{rxNb}%
{%
  name={$L$},
  text={L},
  description={\acs{rx} channel number (Sec.~\ref{sec:analysis:IF signal SIMO array})\textttCom{(rxNb)}},
}

\newglossaryentry{rxIdx}%
{%
  name={$l$},
  text={l},
  description={\acs{rx} channel index (Sec.~\ref{subsec:analysis:Received RF signal})\textttCom{(rxIdx)}},
}


\newglossaryentry{rangeMax}
{%
  name={$\gls{range}{}_{\text{max}}$},
  text={\gls{range}{}_{\text{max}}},
  description={Max. resolvable range (m) \eqref{eq:max range}\textttCom{(rangeMax)}},
}

\newglossaryentry{dopplerMax}
{%
  name={$\gls{doppler}{}_{\text{max}}$},
  text={\gls{doppler}{}_{\text{max}}},
  description={Max. unambiguous velocity ($\text{m}\cdot\text{s}^{-1}$) \eqref{eq:max unambi vel} \textttCom{(dopplerMax)}},
}

\newglossaryentry{rangeRes}
{%
  name={$\Delta \gls{range}$},
  text={\Delta \gls{range}},
  description={Range resolution (m) \eqref{eq:pulse range res} \textttCom{(rangeRes)}},
}

\newglossaryentry{dopplerRes}
{%
  name={$\Delta \gls{doppler}$},
  text={\Delta \gls{doppler}},
  description={Velocity (Doppler) resolution ($\text{m}\cdot\text{s}^{-1}$) \eqref{sec:analysis:Velocity resolution} \textttCom{(dopplerRes)}},
}


\newglossaryentry{c0}
{%
  name={$c_0$},
  text={c_0},
  description={Speed of light ($\text{m}\cdot\text{s}^{-1}$) (Sec.~\ref{sec:analysis:radar equation})\textttCom{(c0)}},
}

\newglossaryentry{f0}%
{%
  name={$f_0$},
  text={f_0},
  description={Start frequency ($\text{s}^{-1}$) \eqref{eq:lfm}\textttCom{(f0)}},
}

\newglossaryentry{wl}%
{%
  name={$\lambda$},
  text={\lambda},
  description={Wavelength (m) (Sec.~\ref{sec:analysis:radar equation}) \textttCom{(wl)}},
}

\newglossaryentry{k_slope}
{%
  name={$k_\text{r}$},
  text={k_\text{r}},
  description={Ramp slope ($\text{s}^{-2}$) \eqref{eq:ramp slope}\textttCom{(k\_slope)}},
}

\newglossaryentry{BW}
{%
  name={$B$},
  text={B},
  description={Signal bandwidth ($\text{s}^{-1}$) \eqref{eq:ramp slope}\textttCom{(BW)}},
}

\newglossaryentry{T_r}
{%
  name={$T_\text{r}$},
  text={T_\text{r}},
  description={Ramp duration (s) \eqref{eq:ramp slope}\textttCom{(T\_r)}},
}

\newglossaryentry{T_s}
{%
  name={$T_\text{s}$},
  text={T_\text{s}},
  description={\acs{adc} sampling time (s) \eqref{eq:sIF Discretization}\textttCom{(T\_s)}},
}

\newglossaryentry{T_pri}
{%
  name={$T_\text{d}$},
  text={T_\text{d}},
  description={Pulse repetition interval (s) (Sec.~\ref{sec:analysis:Resolving the velocity})\textttCom{(T\_pri)}},
}

\newglossaryentry{T_frame}
{%
  name={$T_\text{frame}$},
  text={T_\text{frame}},
  description={Frame duration (s) \eqref{eq:frame duration}\textttCom{(T\_frame)}},
}

\newglossaryentry{fs}%
{%
  name={$f_\text{s}$},
  text={f_\text{s}},
  description={Sampling frequency ($\text{s}^{-1}$) (Sec.~\ref{subsec:analysis:sampling}) \textttCom{(fs)}},
}

\newglossaryentry{modulated}%
{%
  name={$(\cdot)_\star$},
  text={\star},
  description={Modulated (Sec.~\ref{sec:intro:notation})\textttCom{(modulated)}},
}

\newglossaryentry{startPhase}%
{%
  name={$\varphi_{\text{start}} $},
  text={\varphi_{\text{start}}},
  description={Start phase (rad) \eqref{eq:expending LFM phase(t)} \textttCom{(startPhase)}},
}

\newglossaryentry{angleBias}%
{%
  name={$\psi_{\text{bias}} $},
  text={\psi_{\text{bias}}},
  description={\Ac{doa} bias (rad) \eqref{eq:misalignement} \textttCom{(angleBias)}},
}

\newglossaryentry{chImbalance}%
{%
  name={$\vartheta [\gls{chIdx}]$},
  text={\vartheta},
  description={Channel imbalance (rad) \eqref{eq:imbalanced wavefront} \textttCom{(chImbalance)}},
}

\newglossaryentry{codePhase}%
{%
  name={$\varphi_{\gls{txIdx}} [\gls{rampIdx}]$},
  text={\varphi},
  description={Phase code signal for Tx$\gls{txIdx}$ \eqref{eq:phase code definition} \textttCom{(codePhase)}},
}

\newglossaryentry{pc}%
{%
  name={$c_{\gls{txIdx}} [\gls{rampIdx}]$},
  text={c},
  description={Complex code sequence for Tx$\gls{txIdx}$ \eqref{eq:phase code definition} \textttCom{(pc\{k\})}},
}

\newglossaryentry{pcidx}%
{%
  name={$\gls{pc}_{\text{idx},\gls{txIdx}} [\gls{rampIdx}]$},
  text={\gls{pc}_{\text{idx}}},
  description={Constellation index code sequence \eqref{eq:constellation index code sequence} \textttCom{(pc\{k\})}},
}

\newglossaryentry{phaseErr}%
{%
  name={$\delta_{\gls{txIdx}} [\gls{constelIdx}_{\gls{txIdx}}]$},
  text={\delta},
  description={Constellation error (Sec.~\ref{sec:analysis_mimo:const err model})\textttCom{(phaseErr)}},
}

\newglossaryentry{phaseErrVect}%
{%
  name={$\mg{\delta}_{\gls{txIdx}}$},
  text={\mg{\delta}},
  description={Constellation error vector (Sec.~\ref{sec:analysis_mimo:const err model})\textttCom{(phaseErrVect)}},
}

\newglossaryentry{constel}%
{%
  name={$C_{\gls{txIdx}} [\gls{constelIdx}_{\gls{txIdx}}]$},
  text={C},
  description={\ac{tx} constellation (Sec.~\ref{analysis_mimo:subsec:phase-shifter constellation})\textttCom{(constel)}},
}

\newglossaryentry{constelVect}%
{%
  name={$\m{\gls{constel}}_{\gls{txIdx}}$},
  text={\m{\gls{constel}}},
  description={\ac{tx} constellation vector (Sec.~\ref{sec:analysis_mimo:particular ddm codes})\textttCom{(constelVect)}},
}


\newglossaryentry{ddmstep}%
{%
  name={$\Delta \theta_{\gls{txIdx}}$},
  text={\Delta \theta},
  description={\acs{ddm} phase step \eqref{eq:ddm phase code 1}\textttCom{(ddmstep)}},
}

\newglossaryentry{pskNb}%
{%
  name={$\mathcal{N}_{\gls{txIdx}}$},
  text={\mathcal{N}},
  description={\acs{psk} order for Tx$\gls{txIdx}$ (Sec.~\ref{analysis_mimo:subsec:phase-shifter constellation})\textttCom{(pskNb)}},
}

\newglossaryentry{constelIdx}%
{%
  name={$u_{\gls{txIdx}}$},
  text={u},
  description={Index of the constellation vector (Sec.~\ref{analysis_mimo:subsec:phase-shifter constellation})\textttCom{(constelIdx)}},
}

\newglossaryentry{rampConstIdx}%
{%
  name={$\gls{rampIdx}_{\gls{constelIdx}{}_{\gls{txIdx}}}$},
  text={\gls{rampIdx}_{\gls{constelIdx}{}_{\gls{txIdx}}}},
  description={Ramp sub-index \eqref{eq:ramp sub-index def}\textttCom{(constelIdx)}},
}

\newglossaryentry{pskFactor}%
{%
  name={$a_{\gls{txIdx}}$},
  text={a},
  description={\acs{psk} constellation factor \eqref{eq: DDM phase step}\textttCom{(pskFactor)}},
}

\newglossaryentry{DDMbinShift}%
{%
  name={$b_{\gls{txIdx}}$},
  text={b},
  description={\acs{ddm} spectral bin shift \eqref{eq:DDM bin shift}\textttCom{(DDMbinShift)}},
}

\newglossaryentry{diracComb}%
{%
  name={$\Sh_N [\gls{rampIdx}]$},
  text={\Sh},
  description={Discrete Dirac comb of period $N$ \eqref{eq:def discrete dirac comb}\textttCom{(diracComb)}},
}

\newglossaryentry{psg}%
{%
  name={$x_{\text{psg} , \gls{targetIdx}, \gls{txIdx}} [\gls{psgIdx}_{\gls{txIdx}}]$},
  text={x_{\text{psg}}},
  description={\Psg \eqref{eq:peak and spur group def}\textttCom{(psg)}},
}

\newglossaryentry{psgIdx}%
{%
  name={$i_{\gls{txIdx}}$},
  text={i},
  description={\Psg index \eqref{eq:peak and spur group def}\textttCom{(psgIdx)}},
}

\newglossaryentry{psgVect}%
{%
  name={$\m{x}_{\gls{txIdx}, \gls{targetIdx},\gls{rxIdx}}$},
  text={\m{x}},
  description={\Psg vector \eqref{eq:psg vector notation} \textttCom{(psgVect)}},
}

\newglossaryentry{psgVectSum}%
{%
  name={$\m{x}_{\Sigma,\gls{targetIdx},\gls{rxIdx}}$},
  text={\m{x}_{\Sigma}},
  description={Multi-\acs{tx} \psg vector \eqref{eq:multi tx example psg generalized} \textttCom{(psgVect)}},
}

\newglossaryentry{interferer}%
{%
  name={$\Gamma_{a,b} [\gls{rampIdx}]$},
  text={\Gamma},
  description={Interference of $b$ on $a$ \eqref{eq:SR PRPM}\textttCom{(interferer)}},
}


\newglossaryentry{pt}
{%
  name={$P_{\text{Tx}}$},
  text={P_{\text{Tx}}},
  description={Transmit power (W) \eqref{eq:radar equation}\textttCom{(P\_t)}},
}

\newglossaryentry{pr}
{%
  name={$P_{\text{Rx}}$},
  text={P_{\text{Rx}}},
  description={Receive power (W) \eqref{eq:snr def}\textttCom{(P\_r)}},
}

\newglossaryentry{pn}
{%
  name={$P_{\text{Noise}}$},
  text={P_{\text{Noise}}},
  description={System's noise power (W) \eqref{eq:radar equation}\textttCom{(P\_r)}},
}

\newglossaryentry{s_t}
{%
  name={$s_{\text{Tx}} (t)$},
  text={s_{\text{Tx}}},
  description={Transmitted signal (V) \eqref{eq:waveform general}\textttCom{(s\_t)}},
}

\newglossaryentry{s_r}
{%
  name={$s_{\text{Rx}} (t)$},
  text={s_{\text{Rx}}},
  description={Received signal (V) \eqref{eq:Rx signal}\textttCom{(s\_r)}},
}

\newglossaryentry{s_lo}
{%
  name={$s_{\text{LO}} (t)$},
  text={s_{\text{LO}}},
  description={Local oscillator signal (V) \eqref{eq:LO signal}\textttCom{(s\_lo)}},
}

\newglossaryentry{finalGain}%
{%
  name={$A$},
  text={A},
  description={Signal's magnitude (V) \eqref{eq:Rx signal}\textttCom{(finalGain)}},
}

\newglossaryentry{gainTx}%
{%
  name={$A_{\text{Tx}}$},
  text={A_{\text{Tx}}},
  description={\ac{tx} signal's magnitude (V) \eqref{eq:waveform general}\textttCom{(gainTx)}},
}

\newglossaryentry{gainLO}%
{%
  name={$A_{\text{LO}}$},
  text={A_{\text{LO}}},
  description={\acs{lo} signal's magnitude (V) \eqref{eq:waveform general}\textttCom{(gainLO)}},
}

\newglossaryentry{gainTarget}%
{%
  name={$A_{\gls{targetIdx}}$},
  text={A_{\gls{targetIdx}}},
  description={Target's signal's magnitude (V) \eqref{eq:sIF}\textttCom{(gainTarget)}},
}

\newglossaryentry{gainIFTarget}%
{%
  name={$A_{\text{IF},\gls{targetIdx}}$},
  text={A_{\text{IF},\gls{targetIdx}}},
  description={Target's \acs{if} signal's magnitude (V) (Sec.~\ref{subsec:analysis:lpf})\textttCom{(gainIFTarget)}},
}

\newglossaryentry{reflectionPhase}%
{%
  name={$\varphi_{\text{refl}, \gls{targetIdx}}$},
  text={\varphi_{\text{refl}}},
  description={Reflecting material phase rotation (rad) \eqref{eq:Rx signal}\textttCom{(reflectionPhase)}},
}

\newglossaryentry{df}%
{%
  name={$\alpha_{\gls{targetIdx}}$},
  text={\alpha},
  description={Damping factor \eqref{eq:damping factor from radar equation}\textttCom{(df)}},
}

\newglossaryentry{antennaGain}%
{%
  name={$G$},
  text={G},
  description={Antenna gain \eqref{eq:radar equation}\textttCom{(G)}},
}

\newglossaryentry{rcs}%
{%
  name={$\sigma_{\gls{targetIdx}}$},
  text={\sigma},
  description={\Acf{rcs} ($\text{m}^{2}$) (Sec.~\ref{sec:analysis:target model})\textttCom{(rcs)}},
}

\newglossaryentry{snr}%
{%
  name={$\chi$},
  text={\chi},
  description={\Acf{snr} \eqref{eq:snr def} \textttCom{(snr)}},
}

\newglossaryentry{stPhase}
{%
  name={$\varphi_{\text{Tx}} (t)$},
  text={\varphi_{\text{Tx}}},
  description={Waveform's phase function \eqref{eq:freq using phase}\textttCom{(stPhase)}},
}

\newglossaryentry{f_t}
{%
  name={$f_{\text{Tx}} (t)$},
  text={f_{\text{Tx}}},
  description={Waveform's frequency function \eqref{eq:freq using phase}\textttCom{(stPhase)}},
}


\newglossaryentry{s_mix}%
{%
  name={$s_{\text{mix}} (t)$},
  text={s_{\text{mix}}},
  description={Signal after the mixer (V) \eqref{eq:mixer}\textttCom{(s\_mix)}},
}

\newglossaryentry{tf}
{%
  name={$h (t)$},
  text={h},
  description={Filter discrete transfer function (Sec.~\ref{subsec:analysis:lpf})\textttCom{(h)}},
}

\newglossaryentry{s_if}
{%
  name={$s_{\text{IF}} (t)$},
  text={s_{\text{IF}}},
  description={Signal after the \acs{lpf} (V) \eqref{eq:sIF}\textttCom{(s\_if)}},
}


\newglossaryentry{psi_r}
{%
  name={$\psi_{\text{R},\gls{targetIdx}}$},
  text={\psi_{\text{R}}},
  description={Range \acs{if} frequency ($\text{s}^{-1}$) \eqref{eq:range freq}\textttCom{(psi\_r)}},
}

\newglossaryentry{psi_d}
{%
  name={$\psi_{\text{D},\gls{targetIdx}}$},
  text={\psi_{\text{D}}},
  description={Doppler frequency shift ($\text{s}^{-1}$) \eqref{eq:doppler freq} \textttCom{(psi\_d)}},
}

\newglossaryentry{psi_a}
{%
  name={$\psi_{\theta,\gls{targetIdx}}$},
  text={\psi_{\theta}},
  description={DoA frequency ($\text{m}^{-1}$) \eqref{eq:angle frequency}\textttCom{(psi\_a)}},
}


\newglossaryentry{psi_ddm}
{%
  name={$\psi_{\text{DDM},\gls{txIdx}}$},
  text={\psi_{\text{DDM}}},
  description={\acs{ddm} \acs{if} frequency shift ($\text{s}^{-1}$) \eqref{eq:DDM frequency 1} \textttCom{(psi\_ddm)}},
}

\newglossaryentry{IFphase}
{%
  name={$\varphi_{\gls{targetIdx}}$},
  text={\varphi},
  description={Phase (rad) (Sec.~\ref{subsec:analysis:lpf})\textttCom{(IFphase)}},
}


\newglossaryentry{wavefrontSig}
{%
  name={$\phi_{\gls{targetIdx}} [\gls{rxIdx}]$},
  text={\phi},
  description={Wavefront phase vector \eqref{eq:wavefront}\textttCom{(wavefrontSig)}}
}

\newglossaryentry{wavefrontSigVx}
{%
  name={$\phi_{\text{Vx}, \gls{targetIdx}} [\gls{rxIdx}]$},
  text={\phi_{\text{Vx}}},
  description={Wavefront phase vector \eqref{eq:MIMO wavefront}\textttCom{(wavefrontSigVx)}}
}


\newglossaryentry{mF}%
{%
  name={$\m{F}_N$},
  text={\m{F}},
  description={\acs{dft} matrix of size $N$ \eqref{eq:DFT vect}\textttCom{(mF)}},
}

\newglossaryentry{I}%
{%
  name={$\m{I}_N$},
  text={\m{I}},
  description={Identity matrix of size $N$ (Sec.~\ref{sec:analysis_mimo:zero stuffed})\textttCom{(I)}},
}

\newglossaryentry{zeros}%
{%
  name={$\m{0}_{n \times m}$},
  text={\m{0}},
  description={Matrix of zeros of size $n \times m$ (Sec.~\ref{sec:analysis_mimo:zero stuffed})\textttCom{(I)}},
}

\newglossaryentry{I_shift}%
{%
  name={${\m{P}_N}$},
  text={{\m{P}}},
  description={Single position shifting matrix of size $N$ \textttCom{(I\_shift)}},
}

\newglossaryentry{I_stuff}%
{%
  name={${\m{S}_N^s}$},
  text={{\m{S}}},
  description={$(s-1)$-zero stuffing matrix \eqref{eq:zero stuffing matrix def}\textttCom{(I\_stuff)}},
}

\newglossaryentry{I_repeat}%
{%
  name={${\m{R}_N^s}$},
  text={{\m{R}}},
  description={$s$ time duplication matrix \eqref{eq:replication matrix def}\textttCom{(I\_stuff)}},
}


\newglossaryentry{w}%
{%
  name={$w [n]$},
  text={w},
  description={General windowing function \eqref{eq:DFT windowed}\textttCom{(w)}},
}

\newglossaryentry{w1}%
{%
  name={$w_1 [\gls{sampleIdx}]$},
  text={w_1},
  description={Windowing function for the 1st dimension \eqref{eq:SR}\textttCom{(w1)}},
}

\newglossaryentry{w2}%
{%
  name={$w_2 [\gls{rampIdx}]$},
  text={w_2},
  description={Windowing function for the 2nd dimension \eqref{eq:SRD}\textttCom{(w2)}},
}

\newglossaryentry{w3}%
{%
  name={$w_3 [\gls{rxIdx}]$},
  text={w_3},
  description={Windowing function for the 2nd dimension \eqref{eq:S_RDA}\textttCom{(w3)}},
}

\newglossaryentry{s_R}%
{%
  name={${s}_{\text{R}} [\gls{rangeIdx},\gls{rampIdx},\gls{rxIdx}]$},
  text={{s}_{\text{R}}},
  description={Range FFT spectrum \eqref{eq:SR}\textttCom{(s\_R)}},
}

\newglossaryentry{s_RD}%
{%
  name={${s}_{\text{RD}} [\gls{rangeIdx},\gls{dopplerIdx},\gls{rxIdx}]$},
  text={{s}_{\text{RD}}},
  description={Range-Doppler \acs{fft} spectrum \eqref{eq:SRD}\textttCom{(s\_RD)}},
}

\newglossaryentry{s_ch}%
{%
  name={${s}_{\phi , \gls{targetIdx}} [\gls{rxIdx}]$},
  text={{s}_{\phi}},
  description={Target's channel signal \eqref{eq:Sch}\textttCom{(s\_ch)}},
}

\newglossaryentry{s_RDA}%
{%
  name={${\m{S}}_{\text{RDA}} [\gls{rangeIdx},\gls{dopplerIdx},\gls{angleIdx}]$},
  text={{\m{S}}_{\text{RDA}}},
  description={Range-Doppler-Angle \acs{fft} spectrum \eqref{eq:S_RDA}\textttCom{(s\_RDA)}},
}

\newglossaryentry{s_A}%
{%
  name={${s}_{\text{A},\gls{targetIdx}} [\gls{angleIdx}]$},
  text={{s}_{\text{A}}},
  description={Angle \acs{fft} spectrum \eqref{eq:S_A}\textttCom{(s\_A)}},
}

\newglossaryentry{s_NCI}%
{%
  name={${\m{S}}_{\text{NCI}} [\gls{rangeIdx},\gls{dopplerIdx}] $},
  text={{\m{S}}_{\text{NCI}}},
  description={\acs{nci} map \eqref{eq:nci}\textttCom{(s\_NCI)}},
}

\newglossaryentry{A_R}%
{%
  name={${A}_{\text{R} , \gls{targetIdx}}$},
  text={{A}_{\text{R}}},
  description={Peak magnitude on the range spectrum \eqref{eq:ideal slow time signal}\textttCom{(A\_R)}},
}

\newglossaryentry{alpha_R}%
{%
  name={$\alpha_{\text{R} ,\gls{txIdx},\gls{targetIdx}} [\gls{rxIdx}]$},
  text={\alpha_{\text{R}}},
  description={Peak complex value on the range spectrum \eqref{eq:ideal slow time signal}\textttCom{(alpha\_R)}},
}

\newglossaryentry{A_RD}%
{%
  name={${A}_{\text{RD} ,\gls{txIdx},\gls{targetIdx}} [\gls{rxIdx}]$},
  text={{A}_{\text{RD}}},
  description={Peak magnitude on the Doppler spectrum \textttCom{(A\_RD)}},
}

\newglossaryentry{alpha_RD}%
{%
  name={$\alpha_{\text{RD} ,\gls{txIdx},\gls{targetIdx}} [\gls{rxIdx}]$},
  text={\alpha_{\text{RD}}},
  description={Peak complex value on the Doppler spectrum \eqref{eq:sum psg result at rsp level 3}\textttCom{(alpha\_RD)}},
}

\newglossaryentry{A_NCI}%
{%
  name={${A}_{\text{NCI} , \gls{targetIdx}}$},
  text={{A}_{\text{NCI}}},
  description={Peak magnitude on the \acs{nci} map \textttCom{(A\_NCI)}},
}

\newglossaryentry{det_map}%
{%
  name={$\m{B}$},
  text={\m{B}},
  description={Boolean detection map (Sec.~\ref{sec:analysis_mimo:Logical mask for decoding})\textttCom{(det\_map)}},
}

\newglossaryentry{ddm_mask}%
{%
  name={$\m{m}$},
  text={\m{m}},
  description={\acs{ddm} matched filter (Sec.~\ref{sec:analysis_mimo:Logical mask for decoding})\textttCom{(ddm\_mask)}},
}




\newglossaryentry{antennaSpacing}%
{%
  name={$d$},
  text={d},
  description={Antenna spacing (m) (Fig.~\ref{fig:2_analysis/fig/antenna array})\textttCom{(antennaSpacing)}},
}

\newglossaryentry{d_tx}%
{%
  name={$d_\text{Tx} [\gls{txIdx}]$},
  text={d_\text{Tx}},
  description={\acs{tx} antenna position (m) (Fig.~\ref{fig:2_analysis_mimo/fig/mimo})\textttCom{(d\_tx)}},
}

\newglossaryentry{d_rx}%
{%
  name={$d_\text{Rx} [\gls{rxIdx}]$},
  text={d_\text{Rx}},
  description={\acs{rx} antenna position (m) (Fig.~\ref{fig:2_analysis/fig/antenna array}) \textttCom{(d\_rx)}},
}

\newglossaryentry{d_vx}%
{%
  name={$d_\text{Vx} [\gls{chIdx}]$},
  text={d_\text{Vx}},
  description={Virtual antenna position (m) \eqref{eq:MIMO def}\textttCom{(d\_vx)}},
}

\newglossaryentry{arrayOpening}%
{%
  name={$D$},
  text={D},
  description={Array opening (m) \eqref{fig:2_analysis/fig/antenna array}\textttCom{(arrayOpening)}},
}

\newglossaryentry{tdmSig}%
{%
  name={$b_{\gls{txIdx}} [\tdmIdx]$},
  text={b},
  description={\acs{tdm} logical signal (Sec.~\ref{analysis_mimo:subsec:tdm})\textttCom{(tdmSig)}},
}

\newglossaryentry{memSize}%
{%
  name={$S$},
  text={S},
  description={Memory size (MB) (Sec.~\ref{sec:analysis_mimo:Comment on the memory})\textttCom{(memSize)}},
}


\newglossaryentry{Io}%
{%
  name={$\text{I}_\text{0}$},
  text={\text{I}_\text{0}},
  description={\Ac{I} offset error (Fig.~\ref{fig:2_analysis_mimo/fig/phase shifter architecture error})\textttCom{(Io)}},
}

\newglossaryentry{Qo}%
{%
  name={$\text{Q}_\text{0}$},
  text={\text{Q}_\text{0}},
  description={\Ac{Q} offset error (Fig.~\ref{fig:2_analysis_mimo/fig/phase shifter architecture error})\textttCom{(Qo)}},
}

\newglossaryentry{quadErr}%
{%
  name={$\beta_{\gls{txIdx}}$},
  text={\beta},
  description={\acs{iq} quadrature phase error (Fig.~\ref{fig:2_analysis_mimo/fig/phase shifter architecture error})\textttCom{(quadErr)}},
}

\newglossaryentry{gainErr}%
{%
  name={$a_{\gls{txIdx}}$},
  text={a},
  description={\acs{iq} gain error (Fig.~\ref{fig:2_analysis_mimo/fig/phase shifter architecture error})\textttCom{(quadErr)}},
}

\newglossaryentry{rmpIdxCstVec}%
{%
  name={$\m{m}_{\gls{constelIdx}{}_{\gls{txIdx}}}$},
  text={\m{m}},
  description={Ramp cluster index (Sec.~\ref{subsec:autofocus:Analysis at RFFT level})\textttCom{(rmpIdxCstVec)}},
}

\newglossaryentry{cluster}%
{%
  name={$\m{a}$},
  text={\m{a}},
  description={Cluster vector \textttCom{(cluster)}},
}

\newglossaryentry{predistVect}%
{%
  name={$\mathfrak{d}_{\gls{constelIdx}{}_{\gls{txIdx}}}^{n}$},
  text={\mathfrak{d}},
  description={Predistortion vector \eqref{eq:closed loop learning rate} \textttCom{(predistVect)}},
}

\newglossaryentry{lr}%
{%
  name={$\eta$},
  text={\eta},
  description={Learning rate \eqref{eq:closed loop learning rate}\textttCom{(lr)}},
}

\newglossaryentry{kappa}%
{%
  name={$\kappa _{\gls{constelIdx}_{\gls{txIdx}}} [\gls{dopplerIdx}]$},
  text={\kappa},
  description={Spectral contribution of $\gls{constelIdx}_{\gls{txIdx}}$  \eqref{eq:fft ddm code dirac}\textttCom{(kappa)}},
}

\newglossaryentry{barycenter}%
{%
  name={$\beta$},
  text={\beta},
  description={Constellation's center (Sec.~\ref{sec:analysis_mimo:The two Tx case})\textttCom{(beta)}},
}

%% file: acronyms.tex



\acrodefplural{rx}[Rxs]{receivers}
\acrodefplural{tx}[Txs]{transmitters}

\begin{acronym}[CORDIC]

	\setlength{\itemsep}{1ex}
	\setlength{\parskip}{0ex}

	\acro{acc}[ACC]{adaptive cruise control}
	\acro{adas}[ADAS]{advanced driver-assistance system}
	\acro{adc}[ADC]{analog-to-digital converter}
	\acro{aeb}[AEB]{automatic emergency braking}
	\acro{aes}[AES]{automatic emergency steering}
	\acro{alc}[ALC]{automated lane change}
	\acro{awgn}[AWGN]{additive white Gaussian noise}
	\acro{bist}[BIST]{built-in self-test}
	\acro{bpsk}[BPSK]{binary phase shift keying}
	\acro{bs}[BS]{beam steering}
	\acro{bsd}[BSD]{blind spot detection}
	\acro{bw}[BW]{bandwidth}
	\acro{cfar}[CFAR]{constant false alarm rate}
	\acro{cdma}[CDMA]{code-division multiple access}
	\acro{ci}[CI]{coherent integration}
	\acro{cmos}[CMOS]{complementary metal oxide semiconductor}
	\acro{cpi}[CPI]{coherent processing interval}
	\acro{cr}[CR]{corner radar}
	\acro{crma}[CRMA]{chirp-rate multiple access}
	\acro{cw}[CW]{continuous wave}
	\acro{dac}[DAC]{digital-to-analog converter}
	\acro{dbf}[DBF]{digital beamforming}
	\acro{ddm}[DDM]{Doppler-division multiplexing}
	\acro{ddma}[DDMA]{Doppler-division multiple access}
	\acro{dc}[DC]{direct current}
	\acro{dft}[DFT]{discrete Fourier transform}
	\acro{dfft}[D-FFT]{Doppler-FFT}
	\acro{doa}[DoA]{direction of arrival}
	\acro{dr}[DR]{dynamic range}
	\acro{dsp}[DSP]{digital signal processor}
	\acro{eol}[EoL]{end-of-line}
	\acro{fdma}[FDMA]{frequency-division multiple access}
	\acro{fft}[FFT]{fast Fourier transform}
	\acro{fmcw}[FMCW]{frequency-modulated continuous-wave}
	\acro{fov}[FoV]{field-of-view}
	\acro{fps}[fps]{frames per second}
	\acro{fr}[FR]{front radar}
	\acro{ftpc}[FT-PC]{fast-time phase-coding}
	\acro{gcd}[GCD]{greatest common divider}
	\acro{had}[HAD]{highly automated driving}
	\acro{hil}[HIL]{hardware-in-the-loop}
	\acro{hpf}[HPF]{high-pass filter}
	\acro{I}[I]{in-phase}
	\acro{ic}[IC]{integrated circuit}
	\acro{idft}[IDFT]{inverse discrete Fourier transform}
	\acro{ifft}[IFFT]{inverse fast Fourier transform}
	\acro{if}[IF]{intermediate frequency}
	\acro{iq}[I/Q]{in-phase and quadrature components}
	\acro{isar}[ISAR]{inverse synthetic aperture radar}
	\acro{ism}[ISM]{industrial, scientific, and medical}
	\acro{kpi}[KPI]{key performance indicator}
	\acro{lca}[LCA]{lane change assist}
	\acro{lfm}[LFM]{linear frequency modulated}
	\acro{lo}[LO]{local oscillator}
	\acro{lpf}[LPF]{low-pass filter}
	\acro{lrr}[LRR]{long-range radar}
	\acro{lut}[LUT]{look-up table}
	\acro{mae}[MAE]{mean absolute error} 
	\acro{mac}[MAC]{multiplication accumulation}
	\acro{mimo}[MIMO]{multiple-input and multiple-output}
	\acro{mmic}[MMIC]{monolithic microwave integrated circuit}
	\acro{mle}[MLE]{maximum likelihood estimator}
	\acro{mrr}[MRR]{mid-range radar}
	\acro{nci}[NCI]{non-coherent integration}
	\acro{nula}[NULA]{non-uniform linear array}
	\acro{oct}[OCT]{on-chip target}
	\acro{oddm}[ODDM]{odd-DDM}
	\acro{oem}[OEM]{original equipment manufacturer}
	\acro{ofdm}[OFDM]{orthogonal frequency-division multiplexing}
	\acro{pdf}[PDF]{probability density function}
	\acro{pg}[PG]{processing gain}
	\acro{pmcw}[PMCW]{phase-modulated continuous-wave}
	\acro{pll}[PLL]{phased locked loop}
	\acro{prpm}[PRPM]{pseudo-random phase modulation}
	\acro{ps}[PS]{phase shifter}
	\acro{psk}[PSK]{phase shift keying}
	\acro{pri}[PRI]{pulse repetition interval}
	\acro{Q}[Q]{quadrature}
	\acro{qpsk}[QPSK]{quadrature phase shift keying}
	\acro{ransac}[RANSAC]{random sample consensus}
	\acro{rest}[REST]{residue estimation and subtraction technique}
	\acro{rcw}[RCW]{rear collision warning}
	\acro{rcs}[RCS]{radar cross section}
	\acro{rd}[RD]{range-Doppler}
	\acro{rdm}[RDM]{range-Doppler map}
	\acro{rdmult}[RDMult]{Range-division multiplexing}
	\acro{rf}[RF]{radio frequency}
	\acro{rms}[RMS]{root mean square} 
	\acro{rr}[RR]{rear radar}
	\acro{rsp}[RSP]{radar signal processing}
	\acro{rtd}[RTD]{round trip delay}
	\acro{rx}[Rx]{receive}
	\acro{rfft}[R-FFT]{range-FFT}
	\acro{sar}[SAR]{synthetic aperture radar}
	\acro{simo}[SIMO]{single-input and multiple-output}
	\acro{siso}[SISO]{single-input and single-output}
	\acro{slam}[SLAM]{simultaneous localization and mapping}
	\acro{sll}[SLL]{sidelobe level}
	\acro{snr}[SNR]{signal-to-noise ratio}
	\acro{sr}[SR]{side radar}
	\acro{srr}[SRR]{short-range radar}
	\acro{stpc}[ST-PC]{slow-time phase-coding}
	\acro{tdm}[TDM]{time-division multiplexing}
	\acro{tdma}[TDMA]{time-division multiple access}
	\acro{tx}[Tx]{transmit}
	\acro{ula}[ULA]{uniform linear array}
	\acro{vea}[VEA]{vehicle exit assist}
	\acro{vga}[VGA]{variable gain amplifier}

\end{acronym}